\documentclass[useAMS,usenatbib,usegraphicx]{mn2e}
\begin{document}

\title{On the mixing rules for astrophysical inhomogeneous grains.}
\author[N.Maron and O. Maron]{N. Maron$^{1}$\thanks{E-mail:
N.Maron@if.uz.zgora.pl; } and O.Maron$^{2}$\\
$^{1}$Institute of Physics, University of Zielona G\'ora,
	    ul. Podg\'orna 50, 65-246 Zielona G\'ora, Poland\\
$^{2}$Institute of Astronomy, University of Zielona G\'ora,
              ul. Lubuska 2, 65-265 Zielona G\'ora, Poland \\}

\date{Received ; accepted }
\pagerange{\pageref{firstpage}--\pageref{lastpage}} \pubyear{2004}

\maketitle

\label{firstpage}

\begin{abstract}
We present the computation of effective refractive coefficients for inhomogeneous two-component grains with 3 kinds of inclusions with ${\rm m_{incl}=3.0+4.0i, 2.0+1.0i, 2.5+0.0001i}$ and a matrix with ${\rm m_m=1.33+0.01i}$ for 11 volume fractions of inclusions from 0\% to 50\% and wavelengths ${\rm\lambda}$=0.5, 1.0, 2.0 and 5.0 ${\rm \mu m}$. The coefficients of extinction for these grains have been computed using a discrete dipole approximation (DDA). Computation of the extinction by the same method for grains composed of a matrix material with randomly embedded inclusions has been carried out for different volume fractions of inclusions. A comparison of extinction coefficients obtained for both models of grain materials allows to choose the best mixing rule for a mixture. In cases of inclusions with ${\rm m_{incl}}$=2.0+1.0i and 2.5+0.0001i the best fit for the whole wavelengths range and volume fractions of inclusions from 0 to 50\% has been obtained for Lichtenecker mixing rule. In case of ${\rm m_{incl}=3.0+4.0i}$ the fit for the whole wavelength range and volume fractions of inclusions from 0 to 50\% is not very significant but the best has been obtained for Hanai rule. For volume fractions of inclusion from 0 to 15\% a very good fit has been obtained for the whole wavelength range for Rayleigh and Maxwell-Garnett mixing rules.
\end{abstract}


\section{Introduction}
The primary goal of the research we present here has been to use
computational electrodynamics methods to choose the best mixing rule for
different materials (dielectric, semiconductor and metal) of inclusions in
dielectric matrix. Light scattering computation for composite particles 
can be made using the discrete dipole approximation (DDA). \citet{vaidya2001} applied this method to calculate extinction and scattering efficiencies for silicate sphere with embedded graphite inclusions and evaluated interstellar extinction. \citet{andersen2003} tested the programs DDSCAT by \citet{draine2000, draine2000b} and MarCoDES by \citet{markel1998}. The program by \citet{markel1998} is much faster than the one by \citet{draine2000, draine2000b} which is 
very time consuming, however, it is ment to be used mainly for calculating the extinction by sparse clusters, whereas for compact clusters the accuracy of this method is very low. For this reason the DDA methods are not used for mass study of inter- or circumstellar extinction. This is why, many researchers still consider various models of grains such as host material in which other materials are embedded (for example \citet{mathis1989} or \citet{maron1989}) or core-mantle (\citet{jones1988}) and multilayer spherical particles (\citet{voshchin1999}). Besides, the use of a particular model compared to observations can provide information about the inhomogeneous grains. \citet{perrin1990} computed cross section of extinction using: 1$^{0}$ DDA method
assuming cubic lattice with elementary cell treated as inclusion of 10 \AA\ for spherical grain
of radius 100 \AA\ and 2$^{0}$ Mie theory for grain with the same radius.
They considered two mixtures: first a matrix of silicate with complex
index of refraction proposed by \citet{draine1985} and inclusions of thaolin
(organic dielectric with imaginary part of refractive index ${\rm k<0.001}$) and the second case adesite containing inclusions of water ice. Both cases concern only a mixture of dielectrics for two values of volume fraction of inclusions in the infrared region. 
In their paper the effective optical constants applied to the 
Mie theory were obtained from the most popular rules in astrophysics:
Maxwell-Garnett's and Bruggeman's. \citet{perrin1990} compared cross sections of extinction obtained by both methods and stated that the application of effective-medium theories (EMT) to the problem of light scattering by inhomogeneous grains is not straightforward. Other authors (e.g. \citet{chylek2000}) compared the DDA with EMT method but only for very limited range of volume fractions of inclusions, wavelengths and materials.

\section{Mixing rules and effective permittivity}
We limited our study to the grains without electric
charge, magnetic suspectibility and considered only two component mixtures
where inclusions were monodisperse.

In our investigation we tested six mixing rules, not only the most popular in
astrophysics such as Maxwell-Garnett's and Bruggeman's. The considered mixing
rules are presented below.

The Maxwell-Garnett's \citep{mg1904} mixing rule, originally, has been applied to
metal particles encapsuled in an insulating matrix and for small filling
factors, now it is used without restriction. The following equation is known as the Maxwell-Garnett formula \citep{bohren1983}:
\begin{equation}
\varepsilon =\varepsilon _{m}+3f\varepsilon _{m}\frac{\varepsilon
_{i}-\varepsilon _{m}}{\varepsilon _{i}+2\varepsilon _{m}-f\left(
\varepsilon _{i}-\varepsilon _{m}\right) }.
\end{equation}
For composite media consisting of spherical particles of both basic
components without the requirement of complete encapsulation \citet{bruggeman1935} has established the following equation \citep{bohren1983}:
\begin{equation}
f\frac{\varepsilon _{i}-\varepsilon }{\varepsilon
_{i}+2\varepsilon }+(1-f)\frac{\varepsilon
_{m}-\varepsilon }{\varepsilon _{m}+2\varepsilon }=0.
\end{equation}
The Bruggeman's equation is symmetrical, i.e. $\varepsilon _{i}$ and $%
\varepsilon _{m}$ can be exchanged. Similar equation, but not for spherical
inclusions was derived by \citet{landauer1952} for conductivity of metallic mixtures.

For spherical inclusions and random oriented ellipsoids \citet{landau1960} and later, along a different way, \citet{looyenga1965} have found the following relation:
\begin{equation}
\varepsilon ^{\frac{1}{3}}=f\mathit{\,}\varepsilon _{i}^{\frac{1}{3}%
}+\left( 1-f\right) \varepsilon _{m}^{\frac{1}{3}}.
\end{equation}
The Hanai-Bruggeman equation was originally derived by \citet{bruggeman1935} for
composite dielectric consisting of a host material with spherical
inclusions. Hanai in 1961 modified the Bruggeman's equation and generalised it
for complex dielectric permittivities. The Hanai-Bruggeman equation is
following (eq.3.58 in \citet{beek1967}):
\begin{equation}
\frac{\varepsilon _{i}-\varepsilon }{\varepsilon _{i}-\varepsilon _{m}}%
\left( \frac{\varepsilon _{m}}{\varepsilon }\right) ^{\frac{1}{3}}=1-f.
\end{equation}
\bigskip
Other mixing law for homogeneous and isotropic systems has been obtained by 
\citet{lichtenecker1926}. The logarithmic mixing rule for a mixture of two components has the form of:
\begin{equation}
\log \varepsilon =f\log \varepsilon _{i}+\left( 1-f\right)\log \varepsilon _{m}.
\end{equation}
\citet{meredith1960} reconsidered Rayleigh's derivation for a cubical
array of spheres $\varepsilon _{i}$ enclosed in a medium $\varepsilon _{m}$ and obtained an equation that appears to be more satisfactory at high
values of $f$. The following equation describes improved Rayleigh's mixing
rule \citep{beek1967}:
\begin{equation}
\varepsilon =\varepsilon _{m}\frac{\frac{2\varepsilon _{m}+\varepsilon _{i}%
}{\varepsilon _{i}-\varepsilon _{m}}+2f-1.227\frac{2\varepsilon
_{m}+\varepsilon _{i}}{4\varepsilon _{m}+3\varepsilon _{i}}f^{\frac{7}{3}%
}-6.399\frac{\varepsilon _{i}-\varepsilon _{m}}{4\varepsilon
_{m}+3\varepsilon _{i}}f^{\frac{10}{3}}}{\frac{2\varepsilon _{m}+\varepsilon
_{i}}{\varepsilon _{i}-\varepsilon _{m}}-f-1.227\frac{2\varepsilon
_{m}+\varepsilon _{i}}{4\varepsilon _{m}+3\varepsilon _{i}}f^{\frac{7}{3}%
}-2.718\frac{\varepsilon _{i}-\varepsilon _{m}}{4\varepsilon
_{m}+3\varepsilon _{i}}f^{\frac{10}{3}}}.
\end{equation}
In all formulas $f$ is the volume fraction of inclusions and $\varepsilon _{m},\,\varepsilon _{i}$ - the complex dielectric permittivity of
a matrix and inclusion, respectively. The $\varepsilon $ (without a
subscript) is the complex dielectric permittivity of the mixture.
Generally the applicability of all mixing rules is restricted to low concentration of inclusions. In this paper we have extended the search for the best mixing rule up to 50\% of concentration of inclusions.

\section{Details of calculations}
We examined three different cases of the refractive index for inclusions $m_{incl}=3.0+4.0i, 2.0+1.0i, 2.5+0.0001i$. The chosen refractive indices of inclusions refer to those of materials examined by \citet{draine1993}. The matrix in all cases was dielectric with refractive index $m_m=1.33+0.01i$. Because the refractive indices are not additive quantities they have been changed into complex permittivity relative to free space using the following formulae:
\begin{equation}
\varepsilon'=n^2-k^2
\end{equation}
and
\begin{equation}
\varepsilon''=2nk,
\end{equation}
where: $n$ and $k$ are real and imaginary parts of complex refractive index, $\varepsilon'$ and $\varepsilon''$ are real and imaginary parts of complex permittivity.
Next, we have calculated the complex permittivities for the mixture based on the above mixing rules. The computation has been carried out for volume fractions in the range from 0\% to 50\% with 5\% step.
The obtained complex permittivities have been changed into complex refractive indices using the following relations:
\begin{equation}
n=\sqrt{\frac{\sqrt{\varepsilon'^2+\varepsilon''^2}+\varepsilon'}{2}}
\end{equation}
and
\begin{equation}
k=\sqrt{\frac{\sqrt{\varepsilon'^2+\varepsilon''^2}-\varepsilon'}{2}}.
\end{equation}

The extinction coefficients have been computed using the Discrete Dipole Approximation, first proposed by \citet{purcell1973}. DDA methods divide the particle into numerous polarizable elements of volume. The induced dipole polarizations in these cubes are determined
self-consistently, then properties such as the extinction cross section are determined in
terms of the induced polarization. In our study, \citet{draine1994} approach is used to compute the efficiency factor for extinction. The refraction indices for each kind of mixture and all considered mixing rules were calculated using the FORTRAN routine DDSCAT.5a10 given by \citet{draine2000}.  

The efficiency factors for extinction ${\rm Q_{l,j,p}^{homog}}$ for homogeneous spherical grains (each element of DDA array has the same refractive index calculated for the mixture) have been computed for the following wavelengths: ${\rm \lambda=0.5, 1.0, 2.0, 5.0 ~\mu m}$ and grains with radius ${\rm r=0.15 ~\mu m}$ containing 1791 dipoles. The location of dipoles in the array for such pseudospherical particle has been obtained using the routine {\it calltarget.f} also made available by \citet{draine2000}. The subscript p in the symbol ${\rm Q_{l,j,p}^{homog}}$ corresponds to the appropriate mixing rule, j - the volume fraction of inclusion and the subscript l corresponds to wavelength. The determined efficiency factors for extinction have been compared to those computed with the exact Mie theory. It has been found that the differences between the two methods are very small similar to \citet{draine1993} results.

We have applied a mixture of DDA elements with different refractive indices while the ratio of number of DDA elements with refractive index of inclusion to all elements in the grain equals the volume fraction of inclusion.
\begin{figure*}
   \centering
\includegraphics[angle=-90,width=12cm]{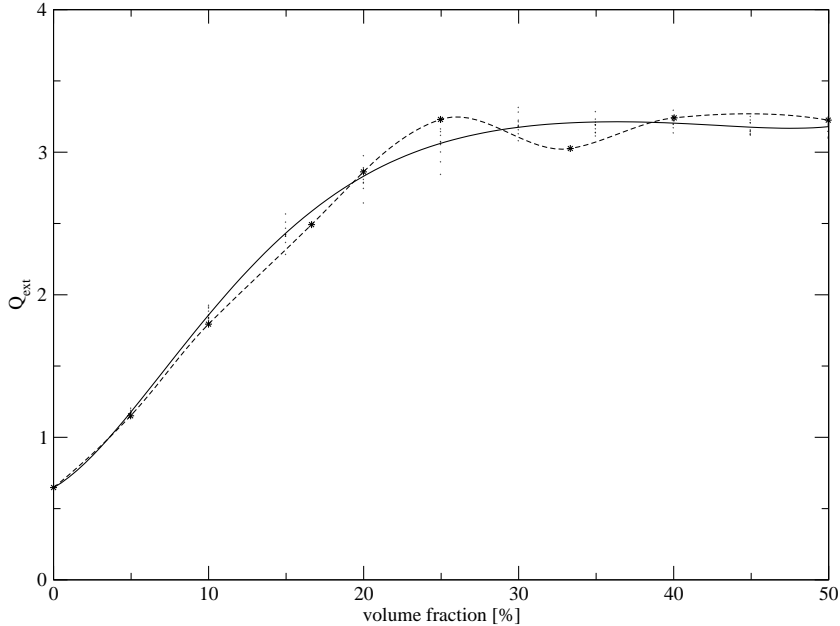}
   \caption{Example of different values of extinction coefficients for different random locations of inclusions with respect to volume fractions of inclusions (dots and solid line best fit) and dependence of extinction on volume fraction for regular location of inclusions (stars and dashed line) both calculated using the DDA. Refractive index of inclusion is ${\rm m_{inc}=3.0+4.0i}$}
\label{fig1}
\end{figure*}
The assumed size of inclusion equals the volume of the DDA element (the size of inclusion cannot be smaller than the element volume considered in DDA approximation). \citet{draine1985} considered how surface granularity affects the accuracy of calculations. In our study validity criteria considered by \citet{draine1985} are satisfied and furthermore the influence of surface granularity is reduced because it is almost the same in both cases: for elements with mixture of materials and for random mixture of elements with two kinds of materials. Similarly, the departure from sphericity for both pseudospheres is the same and does not influence the difference of extinction in both cases. The arrangement of inclusions (DDA elements) is random. In order to obtain random number of DDA elements in the discrete dipole arrays we have used random number generator "Research Randomizer" available at {\it http://www.randomizer.org}. We have generated 10 series of numbers corresponding to the given volume fractions of inclusions out of all 1791 dipoles. The generated numbers have been sorted in ascending order. The location of dipoles in the array for the spherical particle obtained from the routine {\it calltarget.f} was the same as for homogeneous grains with the only difference that for the randomly generated numbers the DDA elements had the refractive index of inclusions and for the others the refractive index of a matrix. Figure~\ref{fig1} presents the dependence of efficiency factors for extinction on the volume fraction for inclusions with regular and random locations. One can see from this figure that the geometrical structure of inhomogeneous grain plays a very important role. Authors of the considered mixing rules have assumed a statistical distribution of inclusions. Therefore, in our calculations the random distributions have been used. Because there is a scattering of results for different random distributions the efficiency factors for extinction have been calculated for 10 different distributions and then averaged. The dependence curves of extinction on volume fractions of inclusions have been interpolated by 5th degree polynomials. For grains with radii of ${\rm r=0.15 \mu m}$ 10 values of efficiency factors for extinction ${\rm Q_{i,j,l}^{rand}}$ depending on random location have been calculated for each of 4 wavelengths $\lambda$=0.5, 1.0, 2.0 and 5.0 ${\rm \mu m}$ by using the computer program DDSCAT.5a10. The subscript i in the symbol ${\rm Q_{i,j,l}^{rand}}$ corresponds to the number of random location of inclusions, j - the volume fraction of inclusion and the subscript l corresponds to wavelength. Next the mean extinction has been calculated as: 
\begin{equation}
Q_{l,j}^{rand}=\frac{1}{10}\sum_{i=1}^{10}Q_{i,j,l}^{rand}.
\end{equation}

\section{Results and discussion}
\begin{figure}
   \centering
      \includegraphics[angle=0,width=84mm]{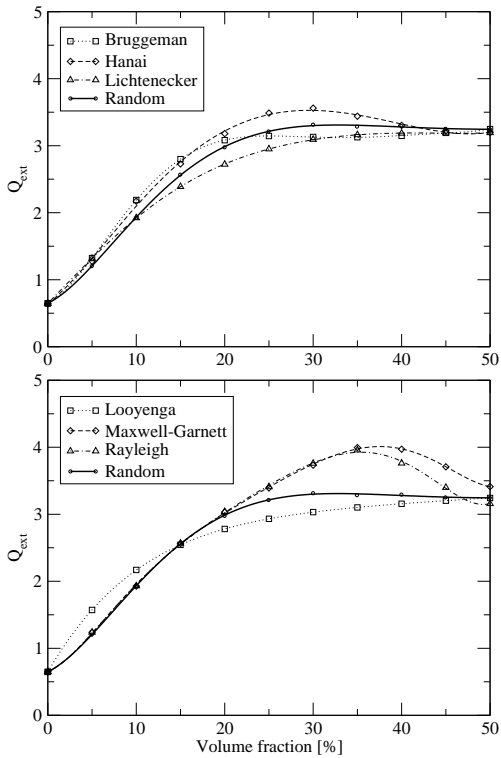}
   \caption{The dependence of efficiency factors for extinction on the volume fraction of inclusions for different mixing rules and wavelengths. Refractive index of the matrix is ${\rm 1.33+0.01i}$ and for inclusion ${\rm m_{inc}=3.0+4.0i}$, the wavelength $\lambda=0.5{\rm \mu}$m. The solid curve "random" shows the best fit for 10 extinction dependencies for randomly distributed DDA elements with refractive index corresponding to inclusion among elements with refractive index of a matrix. The grain radius in all cases is ${\rm 0.15 \mu}$m.}   
\label{fig2}

\end{figure}

\begin{figure}
   \centering
      \includegraphics[angle=0,width=84mm]{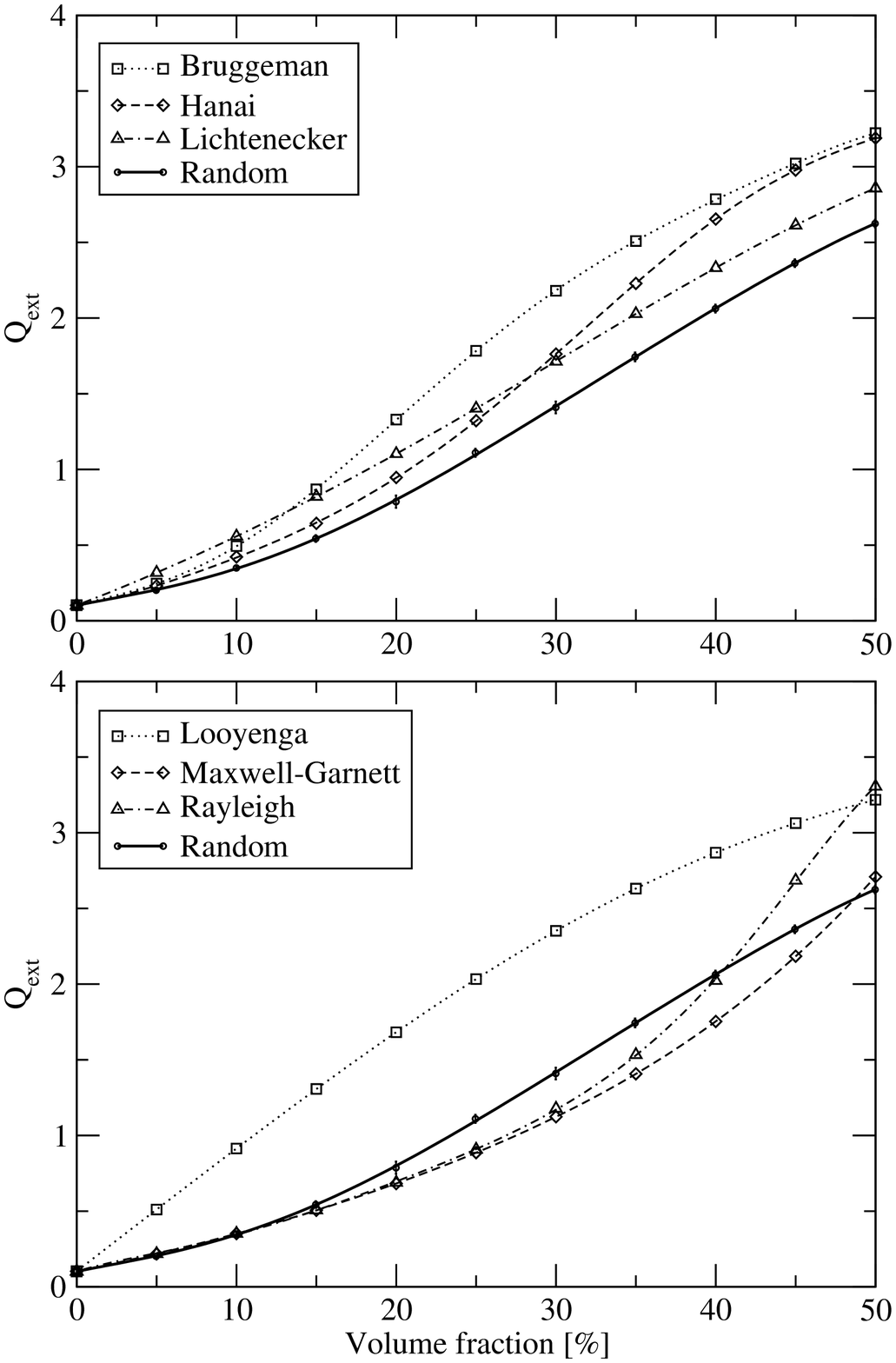}
   \caption{Same as Fig.~\ref{fig2} but for $\lambda=1.0{\rm \mu}$m.}
              \label{fig3}
\end{figure}

\begin{figure}
   \centering
      \includegraphics[angle=0,width=84mm]{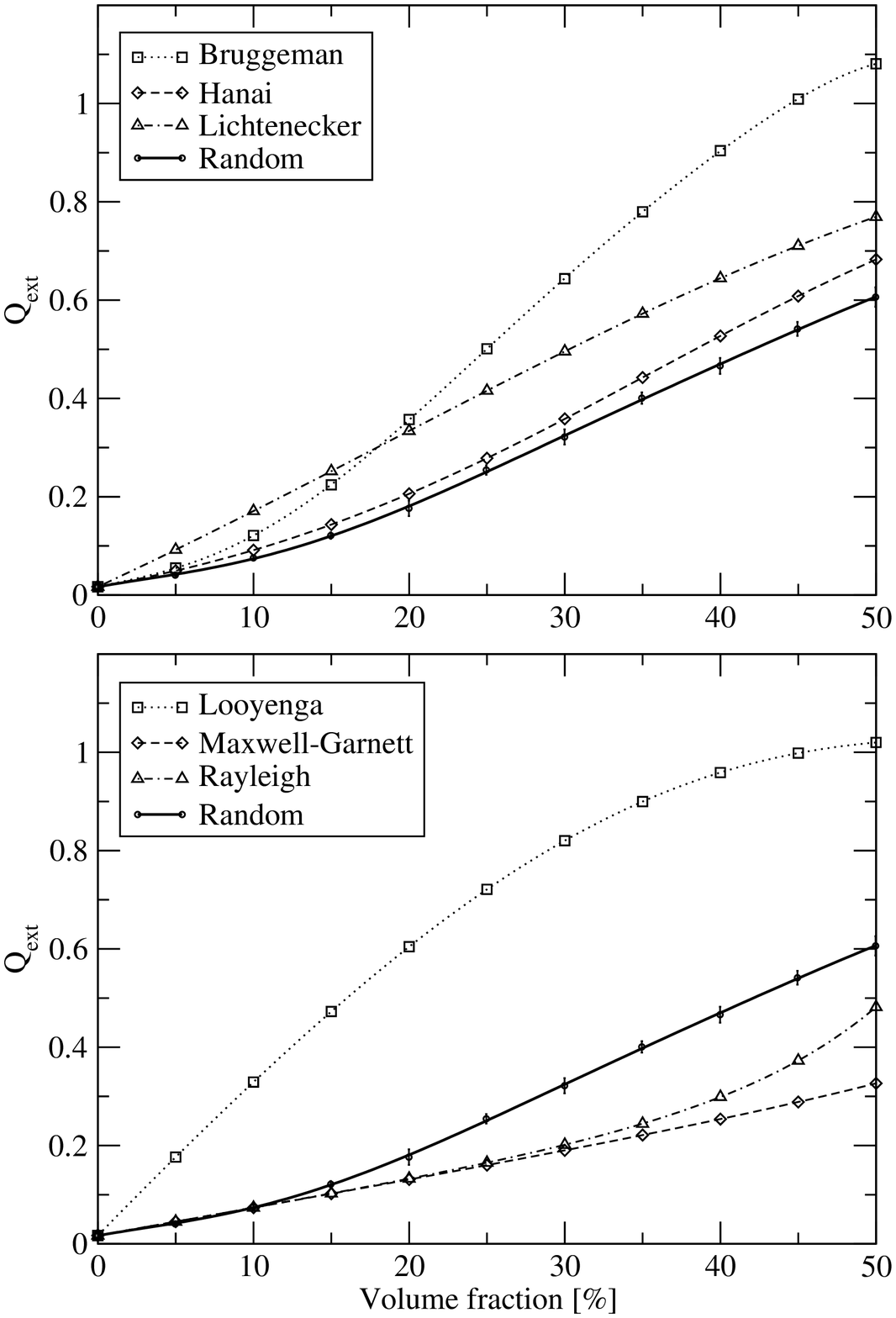}
   \caption{Same as Fig.~\ref{fig2} but for $\lambda=2.0{\rm \mu}$m.}
              \label{fig4}
\end{figure}

\begin{figure}
   \centering
      \includegraphics[angle=0,width=84mm]{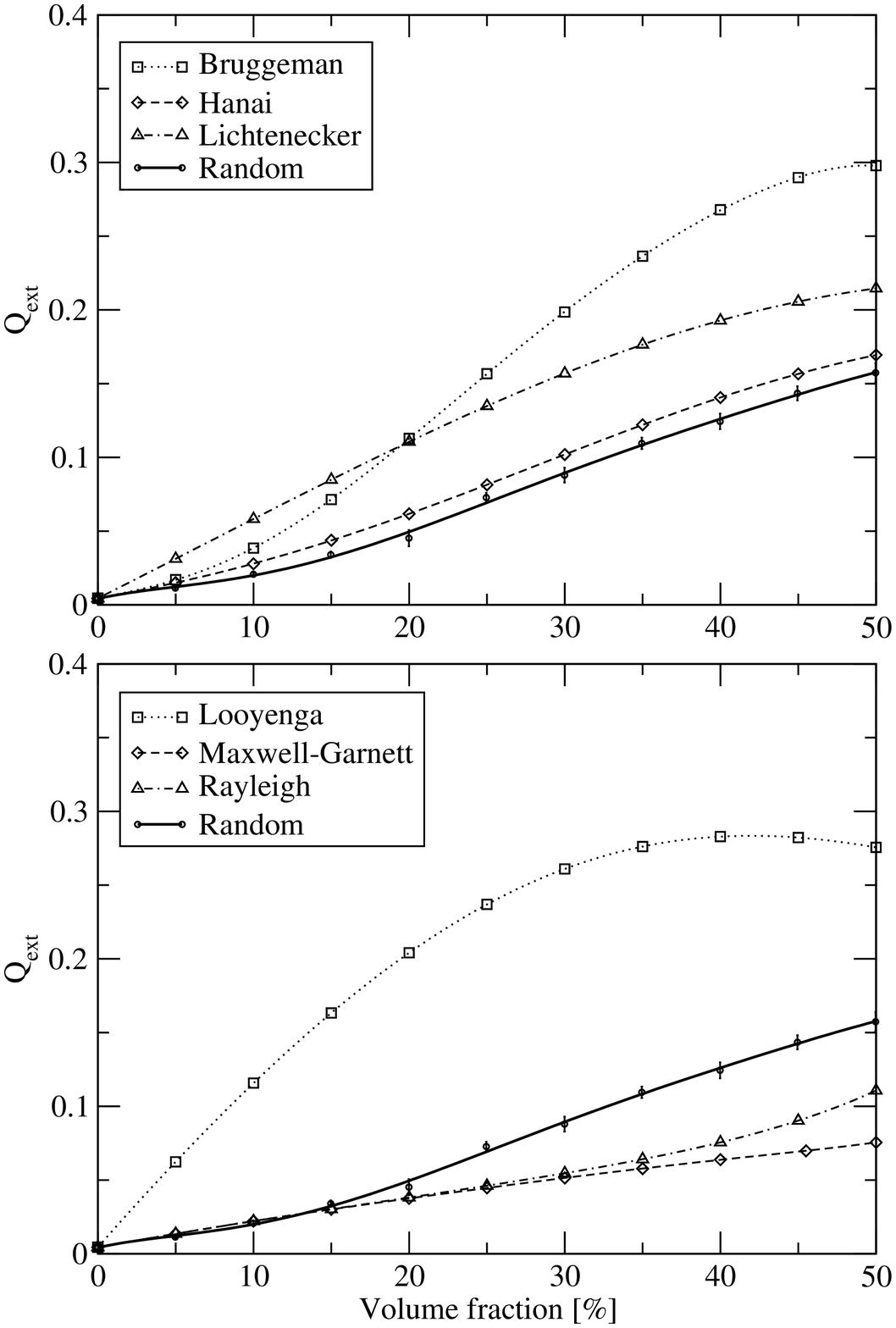}
   \caption{Same as Fig.~\ref{fig2} but for $\lambda=5.0{\rm \mu}$m.}
              \label{fig5}
\end{figure}

\begin{figure}
   \centering
      \includegraphics[angle=0,width=84mm]{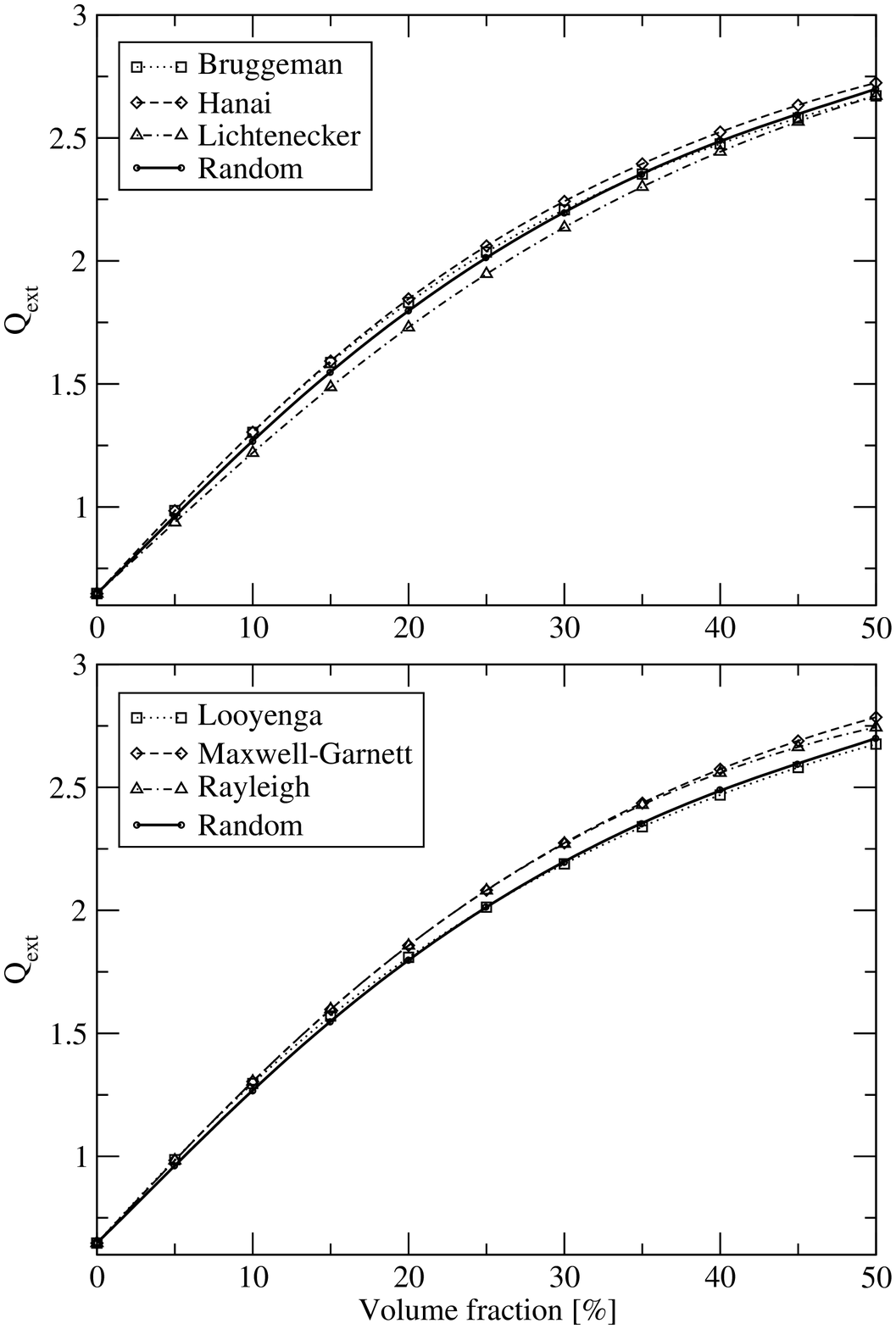}
   \caption{Same as Fig.~\ref{fig2} but for ${\rm m_{inc}=2.0+1.0i}$ and $\lambda=0.5{\rm \mu}$m.}
              \label{fig6}
\end{figure}

\begin{figure}
   \centering
      \includegraphics[angle=0,width=84mm]{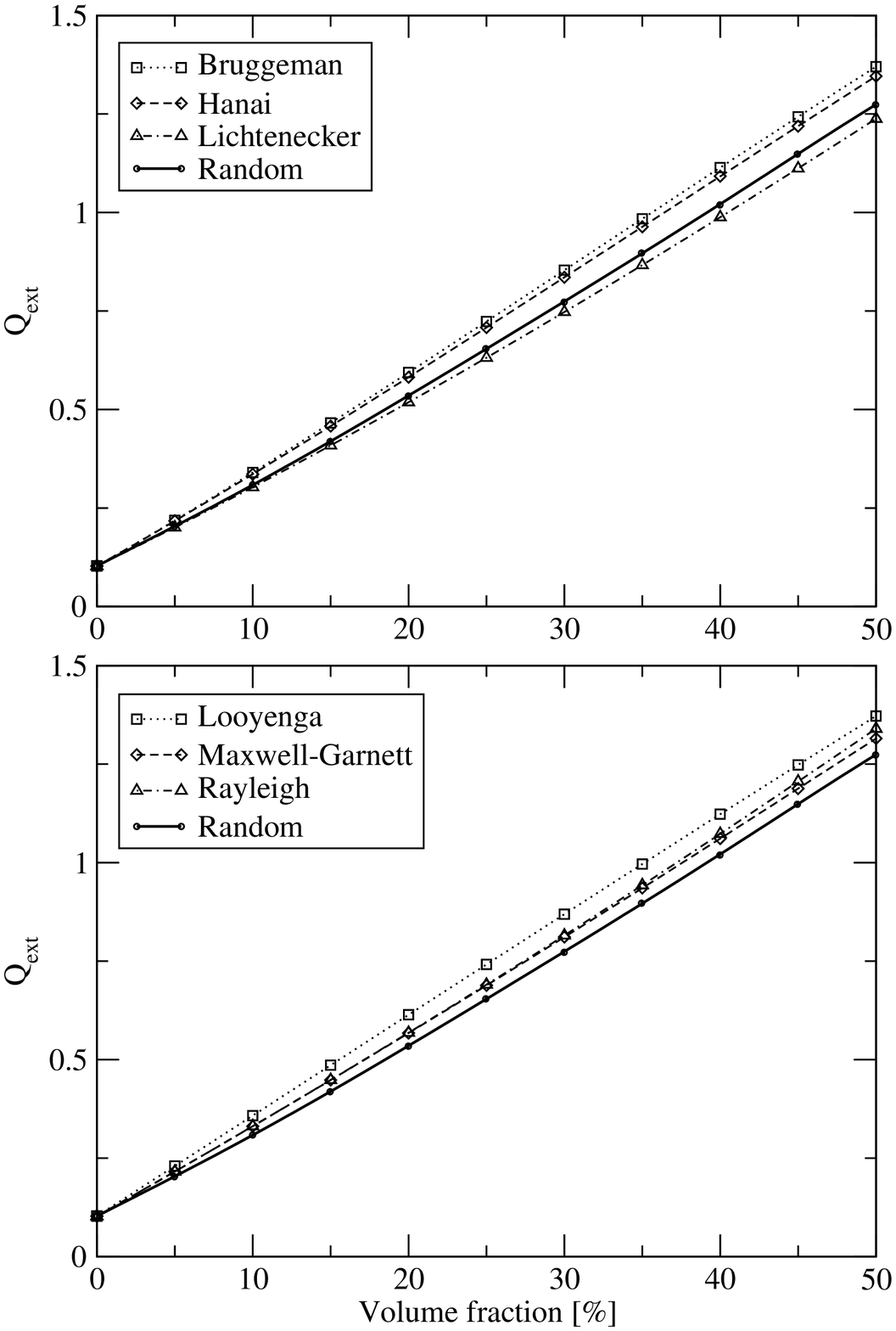}
   \caption{Same as Fig.~\ref{fig2} but for ${\rm m_{inc}=2.0+1.0i}$ and $\lambda=1.0{\rm \mu}$m.}
              \label{fig7}
\end{figure}

\begin{figure}
   \centering
      \includegraphics[angle=0,width=84mm]{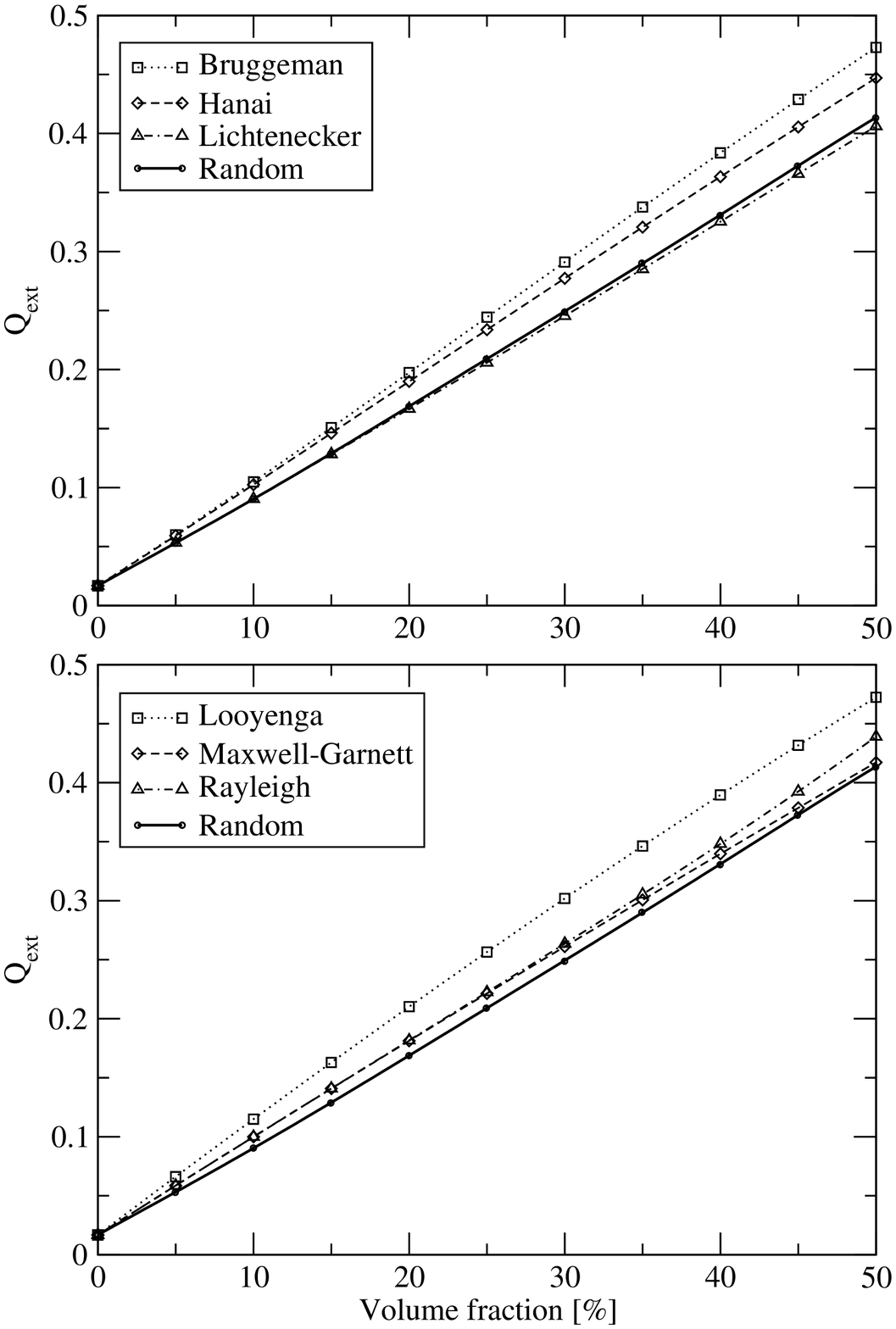}
   \caption{Same as Fig.~\ref{fig2} but for ${\rm m_{inc}=2.0+1.0i}$ and $\lambda=2.0{\rm \mu}$m.}
              \label{fig8}
\end{figure}

\begin{figure}
   \centering
      \includegraphics[angle=0,width=84mm]{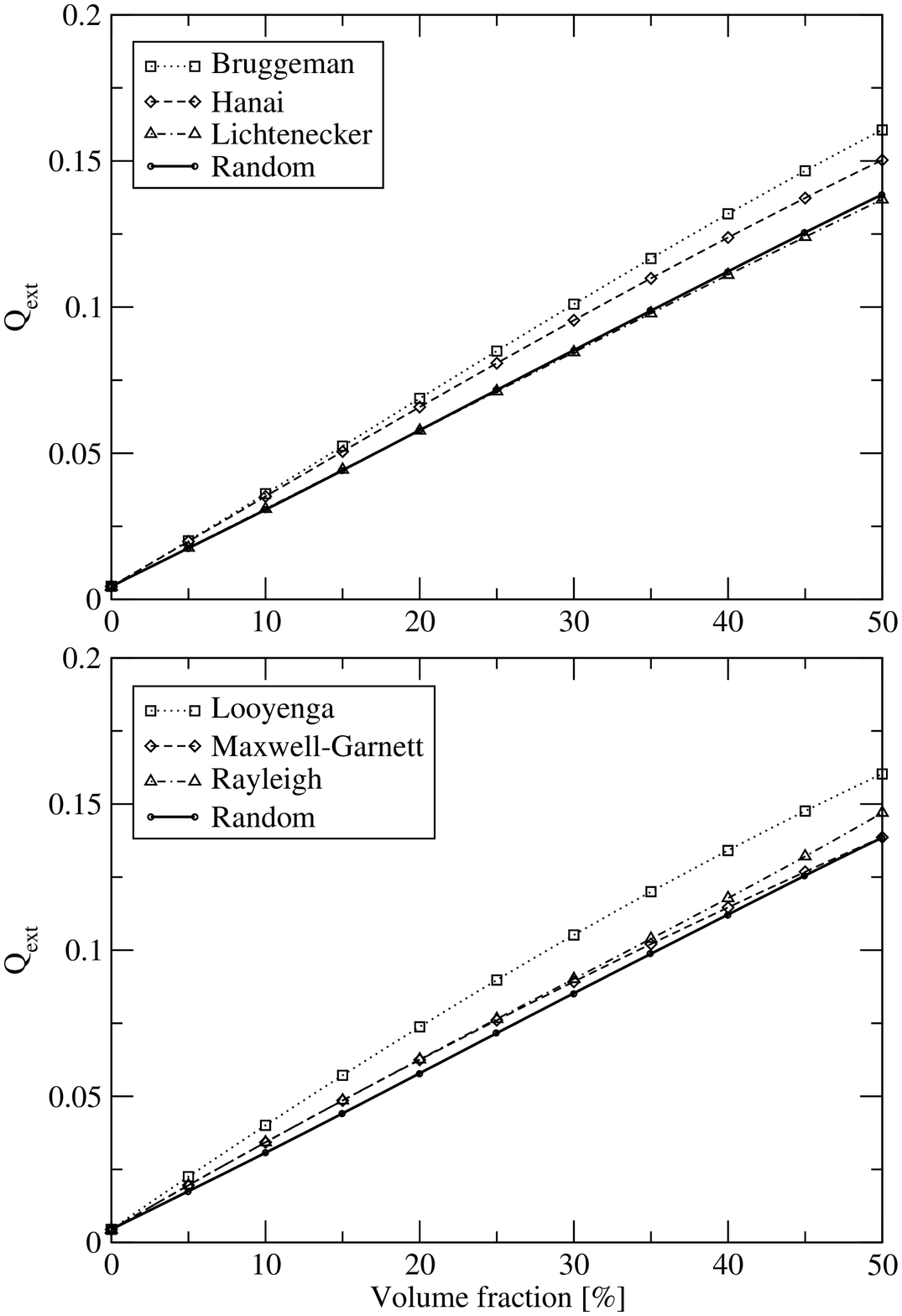}
   \caption{Same as Fig.~\ref{fig2} but for ${\rm m_{inc}=2.0+1.0i}$ and $\lambda=5.0{\rm \mu}$m.}
              \label{fig9}
\end{figure}

\begin{figure}
   \centering
      \includegraphics[angle=0,width=84mm]{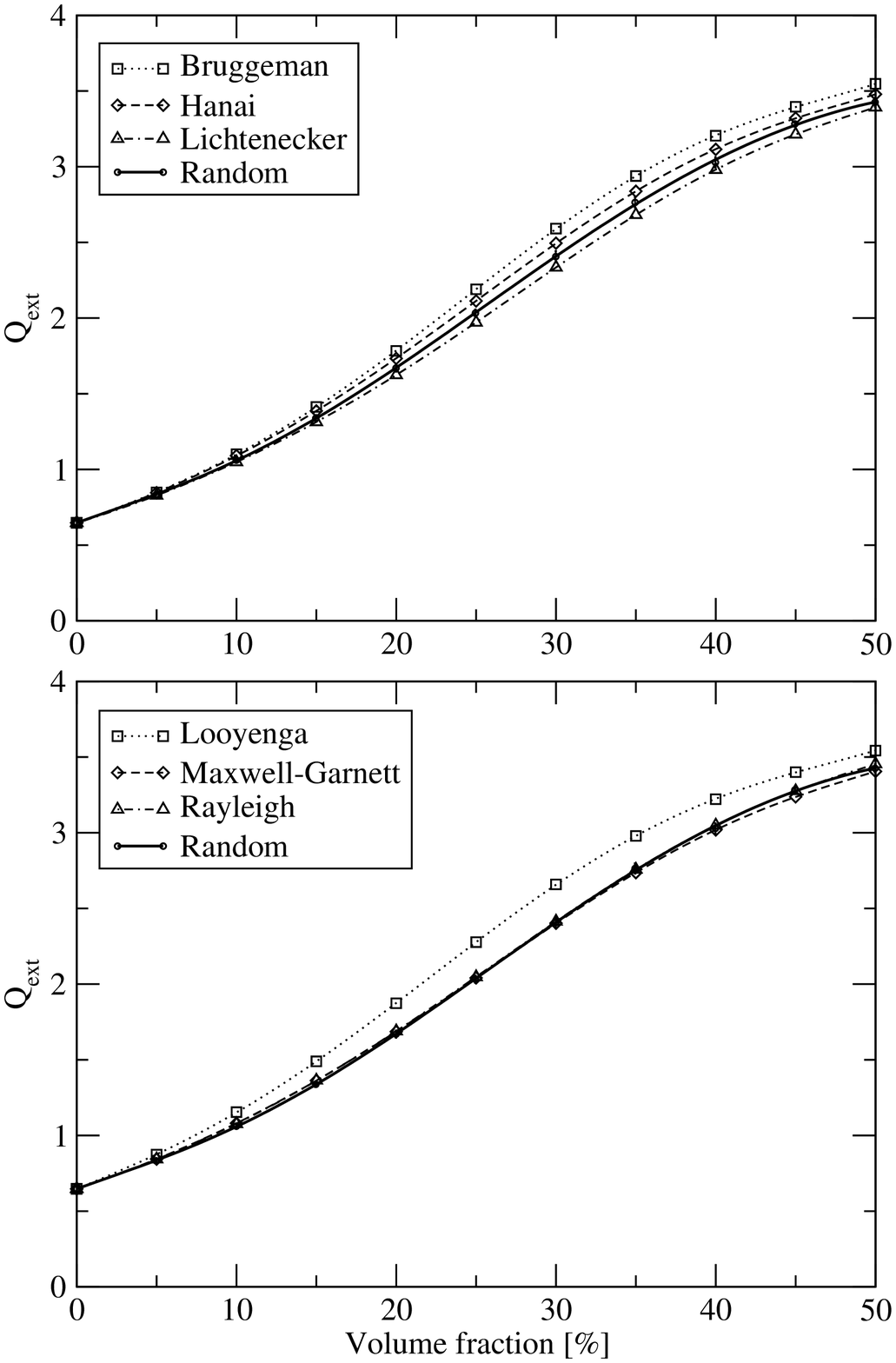}
   \caption{Same as Fig.~\ref{fig2} but for ${\rm m_{inc}=2.5+0.0001i}$ and $\lambda=0.5{\rm \mu}$m.}
              \label{fig10}
\end{figure}

\begin{figure}
   \centering
      \includegraphics[angle=0,width=84mm]{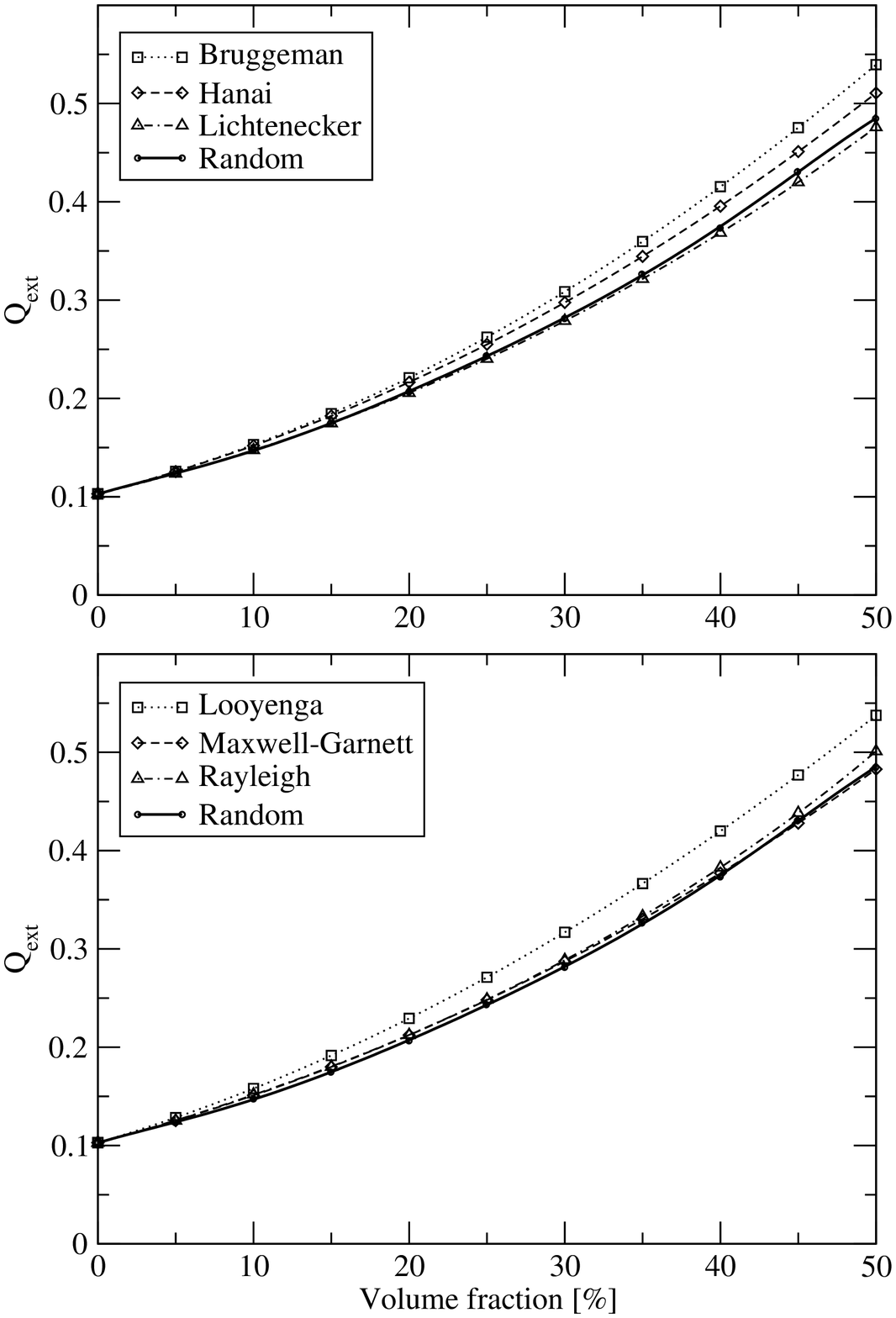}
   \caption{Same as Fig.~\ref{fig2} but for ${\rm m_{inc}=2.5+0.0001i}$ and $\lambda=1.0{\rm \mu}$m.}
              \label{fig11}
\end{figure}

\begin{figure}
   \centering
      \includegraphics[angle=0,width=84mm]{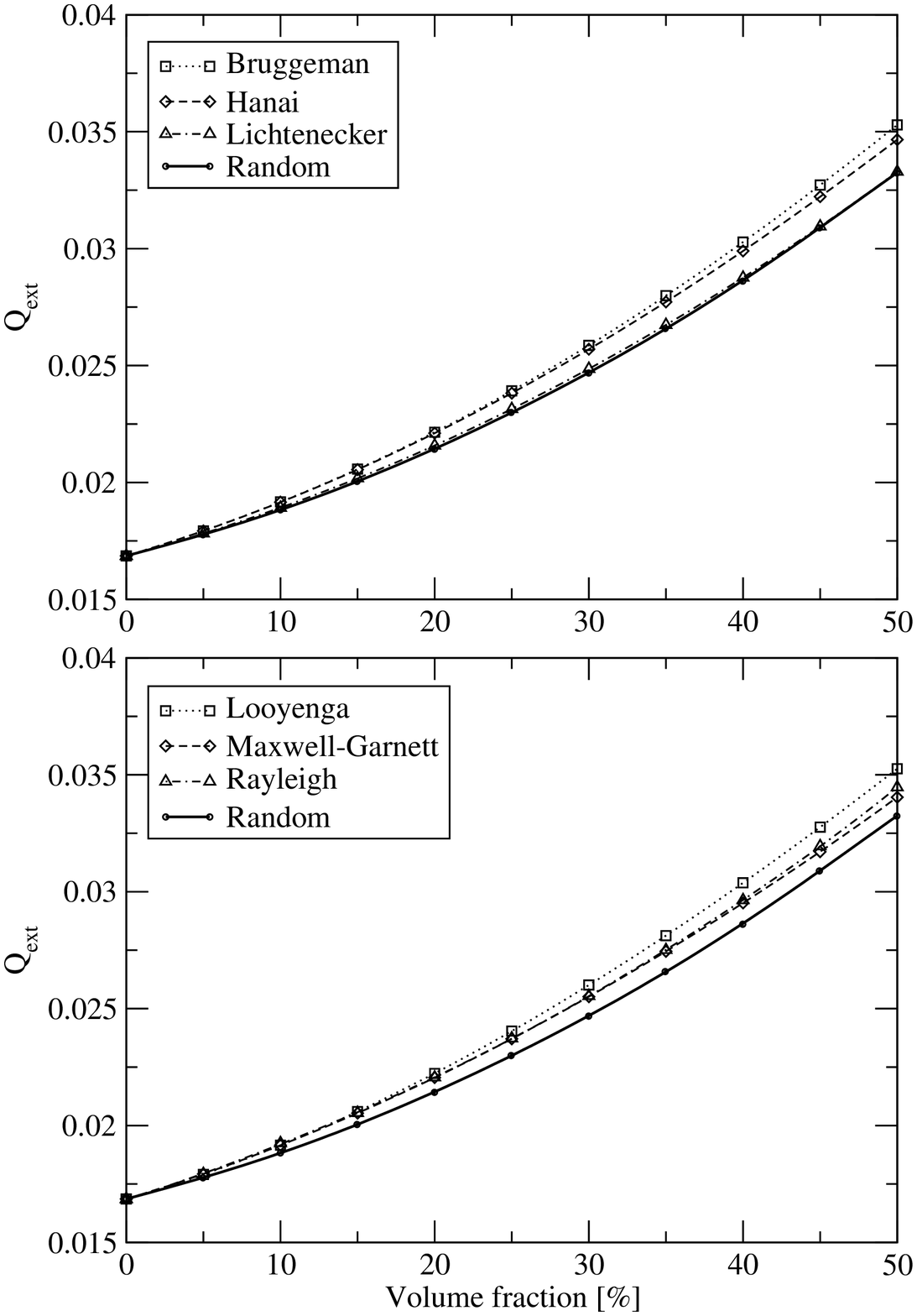}
   \caption{Same as Fig.~\ref{fig2} but for ${\rm m_{inc}=2.5+0.0001i}$ and $\lambda=2.0{\rm \mu}$m.}
              \label{fig12}
\end{figure}

\begin{figure}
   \centering
      \includegraphics[angle=0,width=84mm]{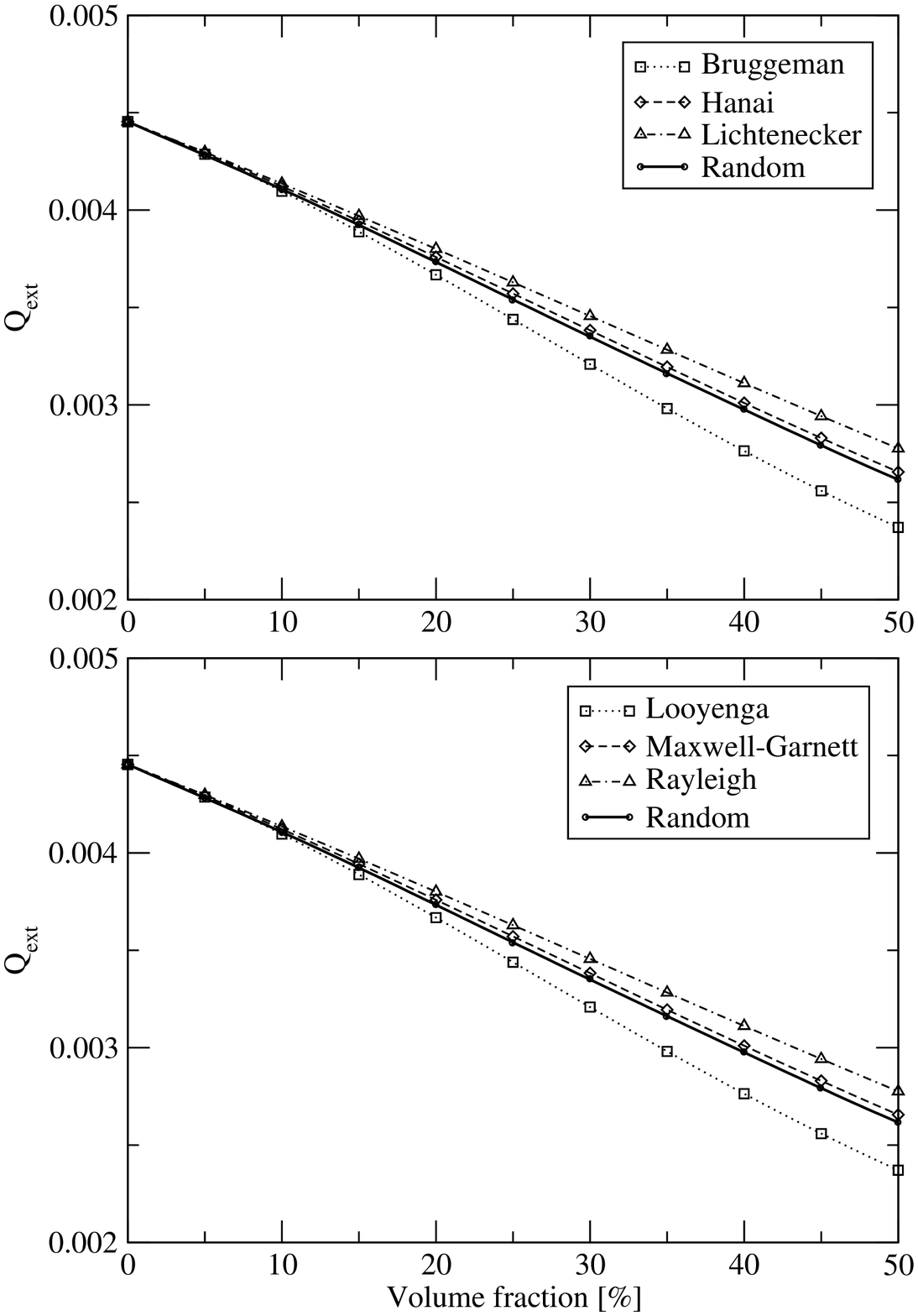}
   \caption{Same as Fig.~\ref{fig2} but for ${\rm m_{inc}=2.5+0.0001i}$ and $\lambda=5.0{\rm \mu}$m.}
              \label{fig13}
\end{figure}

The figures \ref{fig2}-\ref{fig13} show the dependence of mean values of efficiency factors for extinction ${\rm Q_{l,j}^{rand}}$ and ${\rm Q_{l,j,p}^{homog}}$ on the volume fractions of inclusions for the given mixing rule and wavelength and the best fitted curves.

\begin{table*}
\centering
\begin{minipage}{116mm}
\caption[]{Goodness of fit for each mixing rule and each wavelength $\Delta_{l,p}$ - ${\rm m_{inc}=3.0+4.0i}$, ${\rm \sigma=10\%~ of~ Q_{ext}^{rand}}$}
         \label{tab1}
          \begin{tabular}{@{}lccccl@{}}  
            \hline
            
            Mixing rule   & $\Delta_{l,p}(0.5\mu m)$ & $\Delta_{l,p}(1.0\mu m)$ & $\Delta_{l,p}(2.0\mu m)$ & $\Delta_{l,p}(5.0\mu m)$ & $\Delta_{p}$\\
            \hline
            
            Bruggeman & 0.616E$+$00 & 0.643E$-$40 & 0.000E$+$00 & 0.000E$+$00 & 0.000E$+$00 \\
	    Hanai & 0.389E$+$00 & 0.162E$-$06 & 0.911E$-$02 & 0.449E$-$01 & 0.231E$-$06 \\
	    Lichtenecker & 0.996E$+$00 & 0.750E$-$22 & 0.000E$+$00 & 0.000E$+$00 & 0.000E$+$00 \\
	    Looyenga & 0.121E$+$00 & 0.000E$+$00 & 0.000E$+$00 & 0.000E$+$00 & 0.000E$+$00 \\
	    Maxwell$-$Garnett & 0.240E$-$01 & 0.595E$-$01 & 0.535E$-$21 & 0.616E$-$29 & 0.000E$+$00 \\
	    Rayleigh & 0.132E$+$00 & 0.433E$-$01 & 0.120E$-$11 & 0.123E$-$18 & 0.129E$-$26 \\
            
            \hline
         \end{tabular}
\end{minipage}	 
\end{table*}

\begin{table*}
\centering
\begin{minipage}{116mm}
\caption[]{Goodness of fit for each mixing rule and each wavelength $\Delta_{l,p}$ - ${\rm m_{inc}=3.0+4.0i}$, volume fraction of inclusions {\bf 0-15\%}, ${\rm \sigma=10\%~ of~ Q_{ext}^{rand}}$}
         \label{tab2}
          \begin{tabular}{@{}lccccl@{}}  
            \hline
            
            Mixing rule   & $\Delta_{l,p}(0.5\mu m)$ & $\Delta_{l,p}(1.0\mu m)$ & $\Delta_{l,p}(2.0\mu m)$ & $\Delta_{l,p}(5.0\mu m)$ & $\Delta_{p}$\\
            \hline
            
            Bruggeman & 0.826E$+$00 & 0.383E$-$12 & 0.518E$-$34 & 0.000E$+$00 & 0.000E$+$00 \\
	    Hanai & 0.592E$+$00 & 0.182E$-$02 & 0.184E$-$05 & 0.597E$-$09 & 0.950E$-$13 \\
	    Lichtenecker & 0.999E$+$00 & 0.158E$-$37 & 0.000E$+$00 & 0.000E$+$00 & 0.000E$+$00 \\
	    Looyenga & 0.157E$-$09 & 0.000E$+$00 & 0.000E$+$00 & 0.000E$+$00 & 0.000E$+$00 \\
	    Maxwell$-$Garnett & 0.100E$+$00 & 0.997E$+$00 & 0.599E$+$00 & 0.399E$+$00 & 0.994E$+$00 \\
	    Rayleigh & 0.100E$+$00 & 0.997E$+$00 & 0.624E$+$00 & 0.423E$+$00 & 0.996E$+$00 \\
            
            \hline
         \end{tabular}
\end{minipage}	 
\end{table*}

\begin{table*}
\centering
\begin{minipage}{116mm}
\caption[]{Goodness of fit for each mixing rule and each wavelength $\Delta_{l,p}$ - ${\rm m_{inc}=2.0+1.0i}$, ${\rm \sigma=2.5\%~ of~ Q_{ext}^{rand}}$}
         \label{tab3}
          \begin{tabular}{@{}lccccc@{}}  
            \hline
                                 
	    Mixing rule   & $\Delta_{l,p}(0.5\mu m)$ & $\Delta_{l,p}(1.0\mu m)$ & $\Delta_{l,p}(2.0\mu m)$ & $\Delta_{l,p}(5.0\mu m)$ & $\Delta_{p}$\\
            \hline
            
            Bruggeman & 0.929E$+$00 & 0.245E$-$26 & 0.000E$+$00 & 0.000E$+$00 & 0.000E$+$00 \\
	    Hanai & 0.643E$+$00 & 0.216E$-$15 & 0.420E$-$38 & 0.000E$+$00 & 0.000E$+$00 \\
	    Lichtenecker & 0.256E$+$00 & 0.237E$+$00 & 0.987E$+$00 & 0.999E$+$00 & 0.896E$+$00 \\
	    Looyenga & 0.988E$+$00 & 0.000E$+$00 & 0.000E$+$00 & 0.000E$+$00 & 0.000E$+$00 \\
	    Maxwell$-$Garnett & 0.693E$-$01 & 0.103E$-$05 & 0.946E$-$11 & 0.450E$-$13 & 0.317E$-$26 \\
	    Rayleigh & 0.172E$+$00 & 0.158E$-$07 & 0.131E$-$14 & 0.327E$-$17 & 0.736E$-$35 \\
            
            \hline
         \end{tabular}
\end{minipage}	 
\end{table*}

\begin{table*}
\centering
\begin{minipage}{116mm}
\caption[]{Goodness of fit for each mixing rule and each wavelength $\Delta_{l,p}$ - ${\rm m_{inc}=2.5+0.0001i}$, ${\rm \sigma=2.5\%~ of~ Q_{ext}^{rand}}$}
         \label{tab4}
          \begin{tabular}{@{}lccccl@{}}  
            \hline
            
            Mixing rule   & $\Delta_{l,p}(0.5\mu m)$ & $\Delta_{l,p}(1.0\mu m)$ & $\Delta_{l,p}(2.0\mu m)$ & $\Delta_{l,p}(5.0\mu m)$ & $\Delta_{p}$\\
            \hline
            
            Bruggeman & 0.259E$-$06 & 0.371E$-$19 & 0.483E$-$03 & 0.342E$-$05 & 0.143E$-$29 \\
	    Hanai & 0.177E$+$00 & 0.645E$-$04 & 0.277E$-$01 & 0.999E$+$00 & 0.138E$-$02 \\
	    Lichtenecker & 0.682E$+$00 & 0.991E$+$00 & 0.100E$+$01 & 0.367E$-$01 & 0.884E$+$00 \\
	    Looyenga & 0.217E$-$20 & 0.455E$-$32 & 0.110E$-$03 & 0.966E$-$12 & 0.000E$+$00 \\
	    Maxwell$-$Garnett & 0.995E$+$00 & 0.822E$+$00 & 0.307E$+$00 & 0.263E$-$10 & 0.780E$-$05 \\
	    Rayleigh & 0.998E$+$00 & 0.375E$+$00 & 0.132E$+$00 & 0.397E$-$03 & 0.240E$-$01 \\
            
            \hline
         \end{tabular}
\end{minipage}	 
\end{table*}
The best fit of extinction for homogeneous grains consisting of material of averaged refractive index calculated from the mixing rule and extinction of grains consisting of DDA elements of two materials (matrix and inclusions) with different refractive indices (random) was obtained in the following way:\\
a) for the given mixing rule (subscript p) and the given wavelength (subscript l) the $\chi_{p,l}^{2}$ was calculated from
\begin{equation}
\chi_{l,p}^{2}=\frac{1}{10}\sqrt{\sum_{j=1}^{10}\left(\frac{ Q_{l, j}^{rand}-Q_{l,j,p}^{homog}}{\sigma_{l,j,p}}\right)^2},
\end{equation}
where summing is done for volume fractions of inclusions $\rm j$, $Q_{l, j}^{rand}$ is averaged extinction coefficient for randomly located inclusions, $Q_{l,j,p}^{homog}$ is extinction coefficient for homogeneous grains for a given mixing rule. Because the values of $Q_{l, j,i}^{rand}$ for different random locations of inclusions (i) was not treated as equivalent of measured values, therefore $\sigma_{l,j,p}$ is not the standard deviation. $\sigma_{l,j,p}$ is the fraction $\alpha$ of the averaged extinction coefficient $Q_{l, j}^{rand}$.
\begin{equation}
\sigma_{l,j,p}=\alpha \cdot Q_{l,j}^{rand}.
\end{equation}
The calculations of $\chi_{l,p}^{2}$ have been made for ${\rm \alpha=}$0.005, 0.01, 0.025, 0.05 and 0.1. The distribution of $Q_{l, j,i}^{rand}$ values has been assumed to be normal,\\
b) The goodness of fit $\Delta_{l,p}$ has been calculated using the incomplete gamma function ${\rm \Delta_{l,p}=gammq(0.5\nu, 0.5\chi_{l,p}^{2})}$ described in \citet{nr}, where ${\rm \nu=N-M}$ is the number of degrees of freedom, N is the nuber of points in the curve (number of volume fractions of inclusions), M are adjustable parameters (M=l number of mixing rules for calculating the $\chi_{l,p}^{2}$.
The obtained values of fitting coefficients $\Delta_{l,p}$ for each mixing rule and each wavelength for ${\rm \alpha=0.025}$ are presented in Tables~\ref{tab1} and \ref{tab2}, and for the metallic inclusions - ${\rm \alpha=0.1}$ in Tables~\ref{tab3} and \ref{tab4} because for smaller values of ${\rm \alpha}$ the goodness-of-fit coefficients were very small.
For the mixture of given materials determined by a mixing rule different fitting coefficients have been obtained for each wavelength. For different wavelength ranges different mixing rules may be used based on the given values of $\Delta_{l,p}$.\\
c) In order to choose the best mixing rule in the whole considered range of wavelengths the value

\begin{equation}
\chi_{p}^{2}=\sum_{l=1}^{4}\chi_{l,p}^{2}
\end{equation}
has been calculated and the procedure described in b) carried out leading to the obtained goodness-of-fit coefficient $\Delta_{p}$.

Summing up, we have examined 3 different materials as inclusions in a matrix of ${\rm m_{m}=1.33+0.01i}$:
\begin{enumerate}
\renewcommand{\theenumi}{(\arabic{enumi})}
\item {\bf ${\rm \bf m_{inc}=3.0+4.0i}$ (Figures~\ref{fig2}-\ref{fig5})}
\begin{enumerate}
  \item The values of goodness-of-fit coefficients for ${\rm \alpha=}$0.025 and 0.05 are too small to be considered for analysis of the fit, therefore Table~\ref{tab1} shows the values of ${\rm \Delta_{l,p}}$ and ${\rm \Delta_{p}}$ for ${\rm \alpha=}$0.1;
  \item The goodness-of-fit of the extinction curve for each mixing rule strongly depends on the wavelength. For ${\rm \lambda=0.5\mu m}$ the best fit has been obtained for the Lichtenecker rule. However, for the whole range of wavelengths the Hanai rule gives the best fit for this material. It is necessary to point out that the value of the fitting coefficient ${\rm \Delta_{p}}$ is very small. It may be necessary to investigate other mixing rules in order to obtain a better fit;
  \item For a high volume factor of inclusions they may be placed so close that, on one hand, they may create inclusions of much bigger sizes and therefore more vulnerable to skin effect, on the other hand they may exceed the percolation threshold. Therefore, we have considered the fit for the volume factor of inclusions in the range from 0 to 15 \%. The results are presented in Table~\ref{tab2} from which we point out that for the whole wavelength range the best fit is obtained for Rayleigh and Maxwell-Garnett mixing rules. 
\end{enumerate}
\item {\bf ${\rm \bf m_{inc}=2.0+1.0i}$ (Figures~\ref{fig6}-\ref{fig9})}
\begin{enumerate}
   \item The smallest value of ${\rm \alpha}$ which gives the goodness-of-fit coefficients suitable for further analysis is 0.025 and for this value the coefficients ${\rm \Delta_{l,p}}$ and ${\rm \Delta_{p}}$ have been calculated;
   \item For ${\rm \lambda=0.5\mu m}$ the best fit has been obtained for the Looyenga mixing rule (Fig~\ref{fig6}). However, for the whole range of considered wavelengths the Lichtenecker rule gives the best fit (Table~\ref{tab3}).
\end{enumerate}
\item {\bf ${\rm \bf m_{inc}=2.5+0.0001i}$ (Figures~\ref{fig10}-\ref{fig13})}
\begin{enumerate}
   \item The smallest value of ${\rm \alpha}$ which gives the goodness-of-fit coefficients suitable for further analysis is 0.025 and for this value the coefficients ${\rm \Delta_{l,p}}$ and ${\rm \Delta_{p}}$ have been calculated;
   \item For ${\rm \lambda=0.5\mu m}$ the best fit has been obtained for the Rayleigh and Maxwell-Garnett mixing rules. However, for the whole range of considered wavelengths the Lichtenecker rule gives the best fit (Table~\ref{tab4}).
\end{enumerate}
\end{enumerate}

Considering the above results it is possible that different mixing rules should be applied for the same mixture of materials for different wavelengths, especially with higher volume fractions of inclusions.
In spite of their deficiencies like simplifying idealisations while deriving them or basing only upon experiments, the mixing rules are able to provide important information about the inhomogeneous materials.
In astrophysics generally the models by Maxwell-Garnett and Bruggeman are widely used because they are justified by theory. They are based on different topology of inclusions. Other mixing rules, less theoretically justified, are often neglected although they show better agreement with experimental data. However, \citet{zakri} on the basis of effective medium theory found physical grounds for the Lichtenecker rule which, as shown by the calculations, gives better fits for inclusions with refractive indices ${\rm m_{inc}=2.0+1.0i}$ and ${\rm m_{inc}=2.5+0.0001i}$ in the whole considered wavelength range and for ${\rm m_{inc}=3.0+4.0i}$ in the short wavelength region.

\section*{Acknowledgments}

The authors thank the referee for very useful and constructive comments that improved the paper.

\label{lastpage}
\end{document}